# Red-shift effect on the zero field splitting for negatively charged nitrogen-vacancy centers in diamond

Wang Zheng [1, 2], Zhang Jintao [2, *] , Feng Xiaojuan[2]， Xing Li[2]

[1] *Department of Precision Instruments, Tsinghua University, Beijing, China*
[2] *Division of Thermophysics Metrology, National Institute of Metrology, Beijing, China*

The zero field splitting (ZFS) quantifies the energy difference for the ground electron spin-triplet of a nitrogen-vacancy center in the absence of external fields. The values of the ZFS play a key role in determining the Larmor precession of the Bloch sphere and the Rabi oscillation of a spin system. The ZFS is generally detected using coherent spin manipulation by sweeping microwaves (MWs) at frequencies close to resonance with the ZFS. In this letter, we report our experimental observations of the red-shift effect on the ZFS as a function of the MW power for two different thermal environments of a sample. We find an asymptotic property of the red shifts of the ZFS. Given the identical initial thermal equilibrium states of the sample, the differences in the raw values of the ZFS between the two cases randomly vary from 47 kHz to 1505 kHz over the entire experimental range. According to the asymptotic approximation, the differences are reduced to 29-166 kHz with a standard deviation of 49 kHz, suggesting a significant elimination of the red-shift effect. To the best of our knowledge, no other study has addressed the quantification and elimination of the red shift-effect of the MW field dependence using the asymptotic approximation.

A nitrogen-vacancy (NV) color center is a unique point defect in diamond that consists of a substitutional nitrogen atom adjacent to a vacancy within a rigid lattice. The symmetry axis of each NV center is along one of the four (111) crystalline directions. The geometric structure of a NV color center is shown in Fig. 1(a). A negatively charged center ($NV^-$) has a spin-triplet ground state ($S$ = 1), the $|m_s=0\rangle$ and the degenerate $|m_s=\pm1\rangle$, for the zero field splitting (ZFS, denoted by $D$) owing to spin-spin interactions. The ZFS is sensitive to magnetic fields, electric fields, lattice strains and temperatures, allowing a $NV^-$ center to function as a multimodal sensor of a high spatial resolution [1-3]. As shown in Fig. 1(b), the transition of the ground and excitation states occurs at an energy level of 1.945 eV (corresponding to a zero phonon line at a wavelength of 637 nm). Upon the spin-conserving optical excitation, the spin-triplet ground state can be initialized to $|m_s=0\rangle$, because the electrons in $|m_s=\pm1\rangle$ in the excited state more likely undergo intersystem crossing (ISC) to $|m_s=0\rangle$ of the ground state via the associated singlet states [1]. The population in the $|m_s=0\rangle$ state results in higher fluorescence under optical excitation than that in the $|m_s=\pm1\rangle$ state because of spin polarization via ISC. The coherent spin manipulation of the ground spin triplet by sweeping MW frequencies close to resonance with the ZFS can produce prominent photo-luminescence contrast for spin-state-dependent fluorescence. This provides an optical readout, which is referred to as the optically detected magnetic resonance (ODMR) [4]. Several studies have demonstrated the dependence of the ZFS on the temperature and strain of the crystal lattices that host NV centers. Under the zero-strain effect, a $NV^-$ center in diamond acts as a temperature sensor. Acosta *et al.* demonstrated the measurement of temperature by probing the ZFS shift using ODMR [5]. Kusko *et al*. demonstrated a resolution of 0.1 K for a dimension of 200 nm [6]. Toyli *et al.* developed a temperature sensor using a single $NV^-$ center in temperatures of up to 600 K [7]. The high spatial resolution and biocompatibility of $NV^-$ centers in diamond make them promising temperature sensors for living cells [8-14].

* zhangjint@nim.ac.cn

The ZFS is the critical parameter that determines the Larmor precession of a Bloch sphere, and the basis of the ODMR scheme [1]. The Rabi oscillation is based on the resonance with the ZFS. The Ramsey interference and pulsed ODMR (p-ODMR), which are based on the Rabi frequency and the Lamor precession of a Bloch sphere, are mainly used for eliminating the dephasing effect caused by static and gradually varying inhomogeneities, e.g., the dipolar fields of spin impurities in the diamond sample [15]. These dynamic decoupling protocols extend the spin coherence lifetime for increasing phase memory [7,16]. The Rabi frequency must be replaced by the effective Rabi frequency when the ZFS detuning is present [17]. Thus, the sensitivity of spin readout is degraded. Guillebon *et al*. stated that the longitudinal spin-relaxation time of a $NV^-$ center can be decreased by heating a diamond lattice [18]. Such a temperature dependence degrades the detection sensitivity of magnetic fields via quantum lock-in [19].

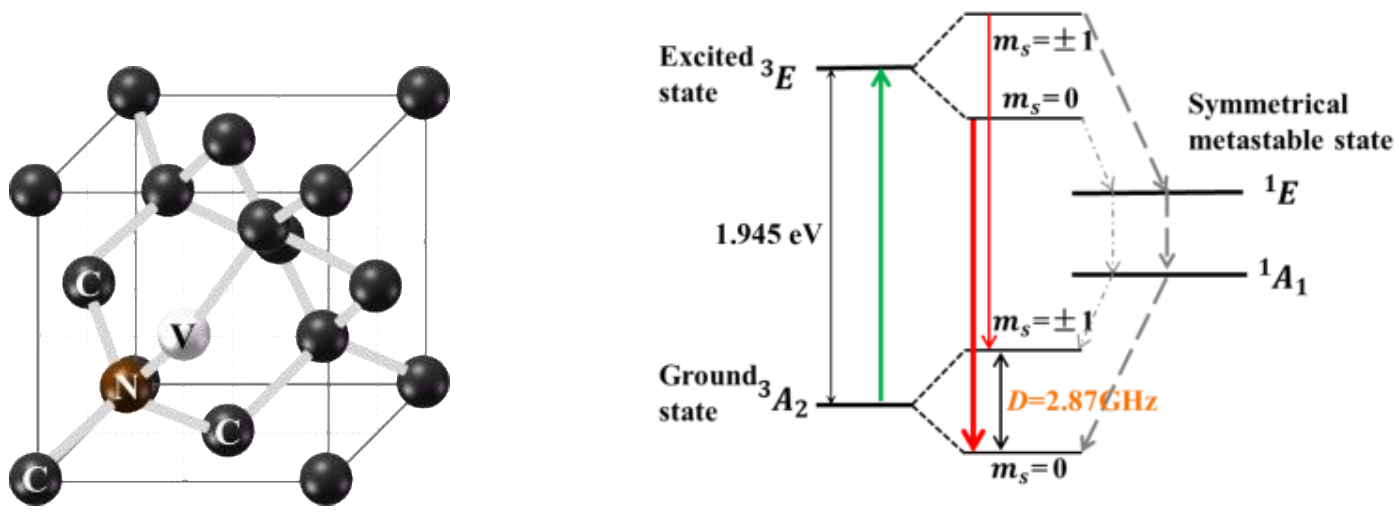

Fig. 1 (a) Geometric structure of NV center. (b) Energy level structure of NV center.

The ISC for the polarization of a spin state via ODMR causes red shifts of the ZFS because the electron-phonon interaction emits phonons, thereby heating a diamond lattice [20]. Our group and others have experimentally observed the laser heating effect on $NV^-$ centers in diamond [21-23]. In addition, the MW irradiation used for manipulating the spin-triplet state heats diamond samples. We previously reported a quantitative investigation of the MW heating effect associated with ODMR [24]. We observed that the MW heating effect, which was stronger than the laser heating effect, depended on the MW power. Fujiwara *et al*. reported that, when measuring the temperatures of biological samples using $NV^-$ centers in diamond, MW irradiation changed the temperature during the measurement process because of MW-induced water heating, resulting in a ZFS shift [25]. Lillie *et al*. detected temperature changes in a sample using a thermistor when only MWs were turned on and observed MW heating [21]. Wang *et al*. discovered that the application of MWs affected the environmental temperatures of diamonds, and quantitatively measured their local temperature changes [26].

Thus, a detailed understanding of the red-shift effect of MW fields on the ZFS is of fundamental importance for the theoretical modeling of the ZFS and metrology using $NV^-$ centers as nanothermometers, magnetometers, etc. Our motivation of this study is to examine the quantification and elimination of the red-shift effect of MW fields on the ZFS. In this letter, we report our experimental study on ensemble $NV^-$ centers driven by the continuous-wave ODMR (cw-ODMR) scheme. The scheme has the advantages of the experimental simplicity and the tolerance of MW inhomogeneities; thus, is widely employed in metrology using $NV^-$ centers. The measurements using this scheme have the desired sensitivities when large ensemble of NV⁻ centers are interrogated with the same optical excitation power [6,11,27,28]. We conduct the study for two different thermal environments using a diamond sample. In Case I, constant current heating is applied to the diamond sample to reach a desired temperature. In Case II, a proportional integral derivative (PID) controller is used to modify the temperature of the diamond sample. We find that the red shifts in the ZFS show asymptotic dependence on MW fields. We demonstrate that the red-shift effect is significantly reduced by extracting the values of the ZFS under a zero MW field through an asymptotic approximation.

The schematic of the experimental setup for the cw-ODMR protocol is shown in Fig. 2(a). The setup is composed of a MW driving facility and a confocal optical path for the photoluminescence of NV⁻ centers and fluorescence collection. A 532 nm laser [MGL-III-532nm, output power > 200 mW, power stability <1% (rms, over 4 h)] is directed towards the diamond sample through the confocal optical path. We use an acousto-optic modulator (AOM, G&H AOM3350-199) as the switch for the laser and an aperture to select the first-order diffraction light modulated by the AOM for the subsequent excitation of the sample. The laser is reflected by a dichroic mirror (DMLP550R) to an objective lens (Olympus, UPLSAPO60XO) and directed towards the sample. The NV⁻ centers undergoing photoluminescence generate fluorescence at approximately 637 nm, which is collected by the confocal optical path, filtered and focused into a multimode fiber for detection by a single-photon counting modulator (SPCM-AQRH-11-FC). The outputs of the detector are acquired using a computer with a data acquisition card (DAQ, PXIe-6363). MWs are generated by Stanford Research System SG386 and transmitted to an amplifier (Mini-Circuits, ZHL-16W-43-S+) via a radio frequency MW switch (Mini-Circuits, ZASWA-2-50DRA+), and finally transmitted to the vicinity of the diamond sample by a 60 μm copper wire antenna laid on a gold-plated printed circuit board (PCB). The PCB board is connected to a 50 Ω impedance resistor for absorbing excess MWs for preventing the reflection signals of MWs from interfering and damaging the instruments. The bulk diamond sample is provided by Element Six, which is fabricated via chemical vapor deposition and has dimensions of 2.6 mm×2.6 mm×0.3 mm. As shown in Fig. 2(b), the sample is bonded to a glass slide attached to the upper surface of the PCB board, and the back surface of the glass slide is attached to a polyimide film heater.

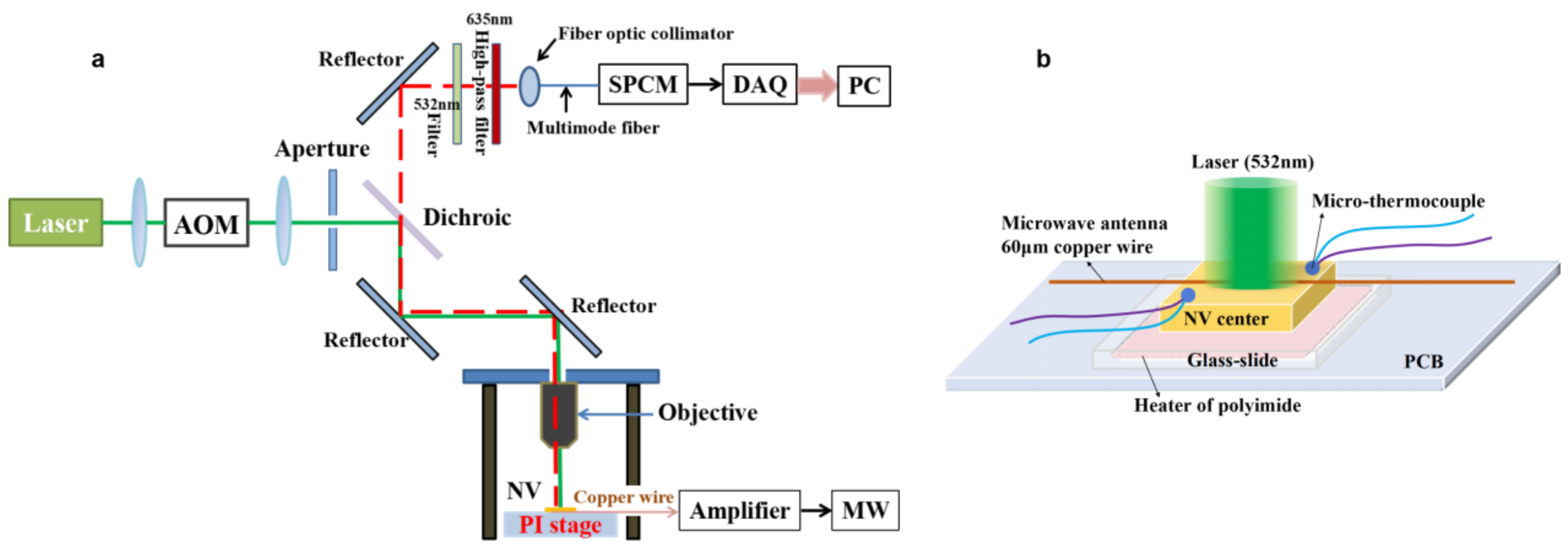

FIG. 2. Schematic of experimental system. (a) Laser illumination system. (b) MW antenna and thermopile.

A thermopile is constructed by connecting two K-type sheathed thermocouples in series. The thermocouple wires have diameters of 0.127 mm. The reference elements of the thermopile are maintained in a cell of the triple point of water (TPW), i.e. 273.16 K. The TPW cell demonstrates a temperature difference of 0.05 mK from the primary standard of the National Institute of Metrology, China. The two probing ends of the thermopile are diagonally attached to the edge of the diamond sample. The diamond sample is wrapped in a thermal insulator.

We use a modified cw-ODMR protocol to suppress the noise generated by the optical excitation and readout [23]. The protocol is illustrated in Fig. 3. The MWs are programmed into a sequence of identical interval pulses, and laser irradiation is applied throughout the measurement. This scheme compares the photoluminescence fluorescence of the $NV^-$ centers with and without MW irradiation to reduce the effect of noise generated by laser irradiation. We select an interval of 5 ms for our measurements. We use a pulse signal generator to synchronize and control MW pulses and fluorescence collection. The MWs are swept in the frequencies from 2850 to 2980 MHz at a step of 0.5 MHz. A MW pulse is applied in 5 ms with a synchronizing photon counting. An extra photon counting is applied in 5 ms in the absence of a MW pulse. The pulsing process is performed 300 times to complete the measurement of a single step of MW sweeping.

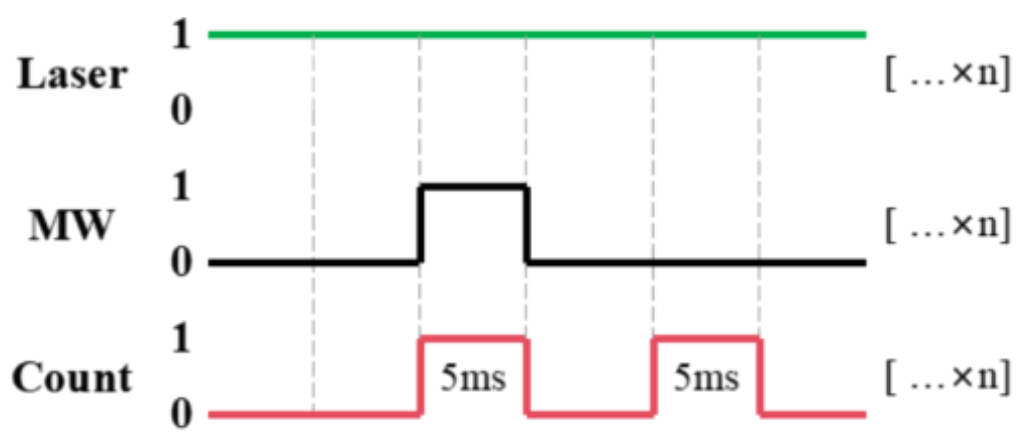

FIG. 3. Time sequences of cw-ODMR.

As stated previously, we design two cases for the thermal environment of the sample. In Case I, we apply a constant current into the film heater to achieve a desired stationary temperature of the diamond sample. The total length obtaining an entire ODMR line is in 160 s. During the period, the temperature fluctuations are maintained within ± 0.05 K. In Case II, we regulate the power of the film heater using the PID controller to achieve a desired temperature. The temperature fluctuations are controlled within ± 0.02 K. We conduct the study in the nominal sample temperatures ranging from 298.15 K to 323.15 K at an interval of 5 K. The MW powers used to manipulate the spin system are -25 dBm, -20 dBm, -15 dBm, -11 dBm, -10 dBm, -5 dBm, 0 dBm, 5 dBm, and 6 dBm. The laser radiation for optical excitation is maintained at a constant power of 3.96 mW.

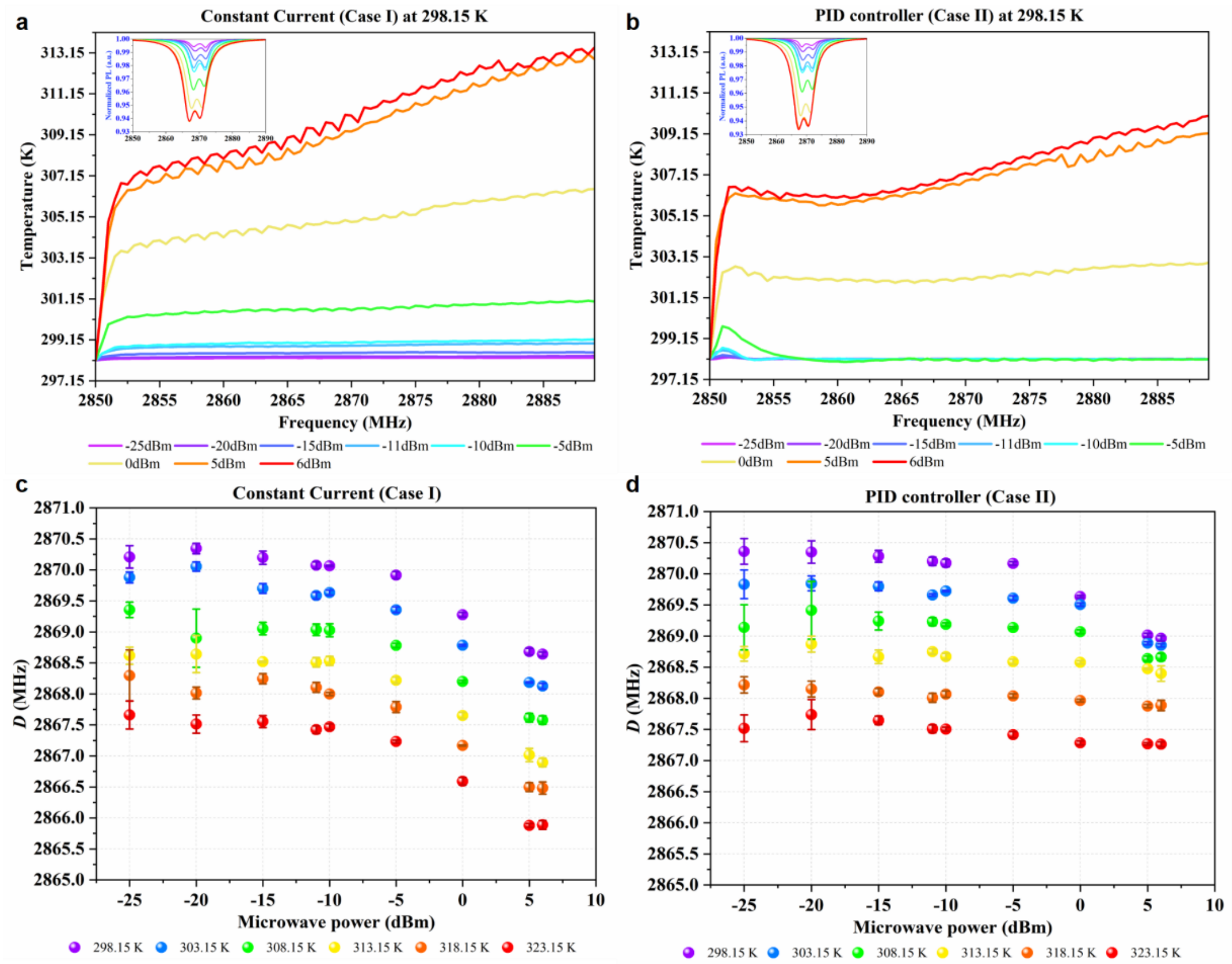

FIG. 4. (a, b) Dependence of diamond sample temperatures on MW field; inset shows the respective ODMR lines. (c, d) Dependence of the red shifts of the ZFS on MW field.

Fig. 4(a) and 4(b) show the MW-power dependence of the sample temperature variations and the ODMR lines for Case I and II; the initial sample temperature is 298.15 K. In Case I, the sample temperature increases with the MW power. The sample temperature rises rapidly during the first 10 s after MW irradiation and gradually thereafter. The minimum temperature rise is 0.13 K at a MW power of -25 dBm and the maximum is 15.4 K at 6 dBm. The ODMR lines exhibit increasing contrast and broadening.

In Case II, there is a threshold MW power for the sample temperature. Below the threshold, the sample temperature rapidly rises after the MW irradiation is applied and then quickly converges in decaying oscillations to the original sample temperature. The sample temperature oscillates around the original temperature with amplitudes of ± 0.05 K. When the MW power exceeds the threshold of 0 dBm for the current study, the PID controller fails to control the sample temperature, which increases with time. However, the magnitude of the increase is less than that for Case I. The ODMR lines are similar to those for Case I. We presented in Supplemental Material the results for other initial sample temperatures [29].

Figs. 4(c) and 4(d) show the measured values of the ZFS as a function of the MW power for Cases I and II, respectively. The ZFS values for both cases exhibit red shifts of the MW-dependence. We examine the red shifts occurring in Case II for MW powers below the threshold. As shown in Fig. 4(b), no apparent temperature rise is recorded by the thermopile attached to the edge of the diamond sample for MW powers below -5 dBm. Our experiment is conducted in an electromagnetically shielded room. No additional electromagnetic field is applied, except for the MW fields, to manipulate the ground electronic spin triplet. The diamond sample is strain free. We argue that the MW heating effect occurs in the diamond sample even though the MW power is below the threshold. The wire antenna produces an inhomogeneous field in the sample. We can infer that the sample central is slightly hotter than the sample edge. The thermopile attached to the edge of the diamond sample does not detect the rise in temperature at the center of the sample. The PID controller acts according to the edge temperature measured by the thermopile. Thus, the temperature measured by the thermopile is lower than that in the central region of the sample. We reported an intensified MW field in the central region observed using an infrared camera [24]. The previous experimental observations support this reasoning.

Acosta *et al*. ascribed the thermal expansion to the temperature dependence of the ZFS [5,30]. Doherty *et al.* stated that a model based on pure thermal expansion did not accurately describe the temperature dependence of the ZFS [31]. The atomic displacement arising from thermal expansion and the lattice vibration were responsible for the temperature dependence of the ZFS. Tang *et al*. reported that atomic thermal expansion negligibly affected the ZFS [32]. The atomic vibrations around the equilibrium positions caused by phonon excitations at a finite temperature dominated the temperature dependence of ZFS shifts. The phonon occupation number affected the phonon effect. Cambria *et al.* inferred that, for the spin triplet of $NV^-$ centers, changes in temperature affected the positions of atoms in the diamond lattice which modulated the electronic wavefunction [33]. The temperature-dependent ZFS shifts were proportional to the occupation numbers of two representative phonon modes. Tang *et al.* and Cambria *et al*. reported that the phonon modes dominated the temperature-dependent ZFS shifts. Gali stated that, as the electron spin density was highly localized to the three carbon atoms adjacent to the vacancy, the ZFS could be accurately modelled by accounting for the temperature-dependent positions of the atoms that were the closest to the defect [34]. Therefore, in this study, we ascribe the observed red shifts to the additional vibration of the diamond lattice thermally excited by the MW fields.

Figs. 4(c) and 4(d) show that a lower MW power results in a smaller red-shift effect in the expense of larger error bars for the values of the ZFS and more inferior contrasts of the ODMR lines. In Case I, the random errors for the measured ZFS values, which are obtained using standard deviations, are 87-413 kHz at a MW power of -25 dBm. In contrast, the measurements performed using a MW power of -5 dBm result in standard deviations of 10-61 kHz for the ZFS values and the red shifts of 294-574 kHz. This implies that a high MW power results in small standard deviations of the ZFS values in the expense of substantial red shifts in the ZFS. A similar phenomenon occurs in Case II. Fig. 4(d) shows that the random errors for the measured ZFS values are 118-364 kHz at the MW power of -25 dBm. In addition, the ZFS values for Case I differ from those for Case II by 47-1505 kHz over the entire experimental range.

We observe that Figs. 4(c) and 4(d) show a similar asymptotic behavior of the MW-power dependence. This phenomenon indicates that the red shifts asymptotically disappear as the MW power decreases. Thus, zero MW irradiation should not affect the ZFS of the $NV^-$ centers in diamond. On this basis, we create an exponential asymptotic fit of the ZFS values related to the MW power. We denote the MW power by $Q$ (dBm), and the ZFS by $D$ (MHz). We consider the MW power as the free variable and obtain the following exponential asymptotic fit:

$$D = a \times 10^{Q/20} + b. \tag{1}$$

We pictured the fits in Fig. 5(a) and 5(b) for Case I and II, respectively. Eq. (1) well fits $D$ and $Q$. Then, we convert $Q$ into $Q_1$ in units of mW ($Q_1=10^{Q/10}$). Table I provides the respective conversion for all MW powers.

TABLE I. Unit conversion of the MW power.

| **$Q$ (dBm)** | **$Q_1$ (mW)** | **$x=Q_1^{1/2}$** |
|---|---|---|
| -25 | 0.003162278 | 0.056234133 |
| -20 | 0.01 | 0.1 |
| -15 | 0.031622777 | 0.177827941 |
| -11 | 0.079432823 | 0.281838293 |
| -10 | 0.1 | 0.316227766 |
| -5 | 0.316227766 | 0.562341325 |
| 0 | 1 | 1 |
| 5 | 3.16227766 | 1.77827941 |
| 6 | 3.981071706 | 1.995262315 |

Assuming $x=Q_1^{1/2}$, the exponential fit is converted into a linear fit below

$$D = ax + b. \tag{2}$$

We pictured the linear fits for Case I and II in Fig. 5(c) and 5(d), respectively. According to the linear fits, the intercept $b$ accounts for the ZFS value at zero MW irradiation, i.e., the unperturbed ZFS value.

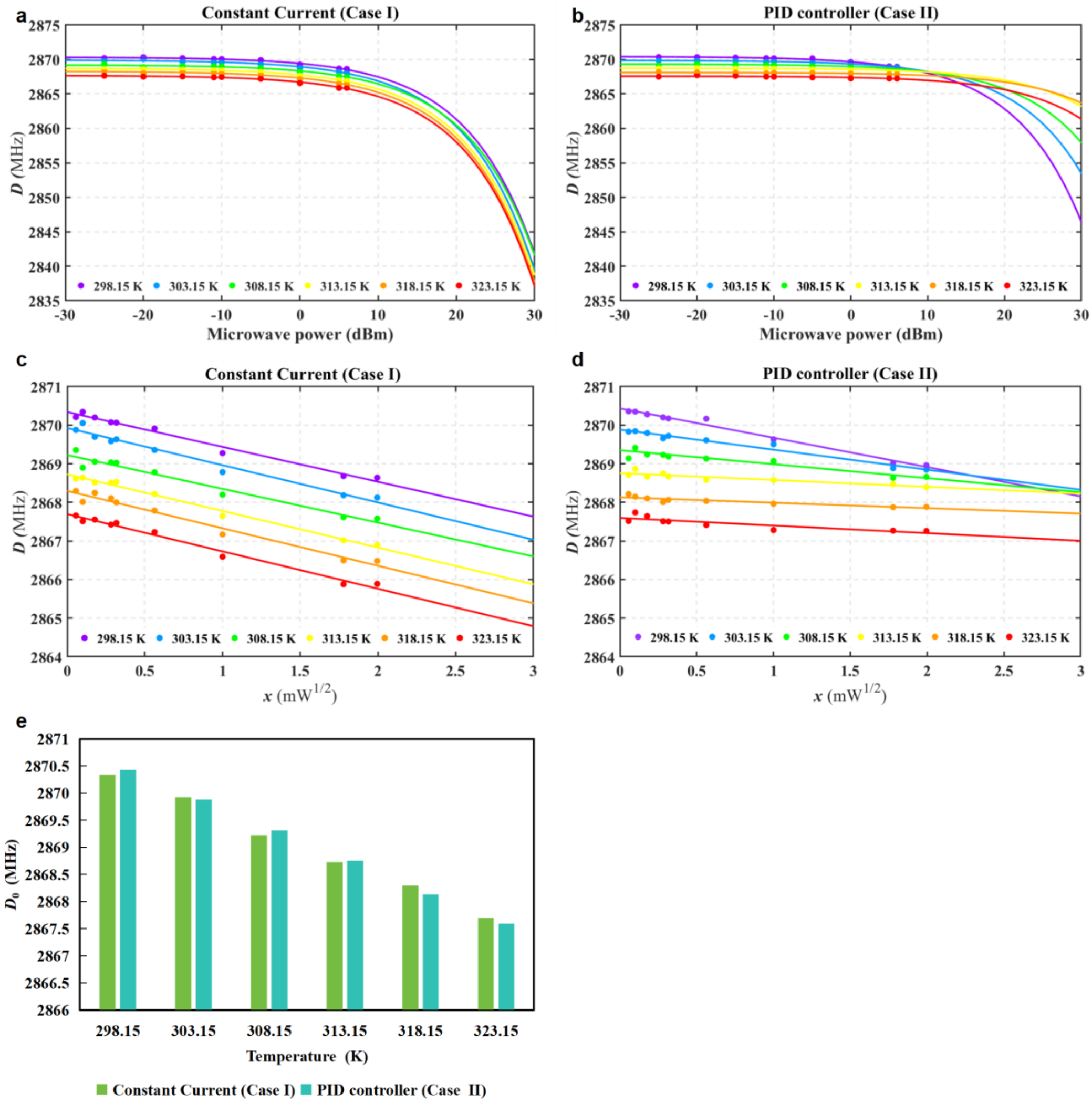

FIG. 5. (a, b) Exponential asymptotic fitting of the ZFS values measured at different temperatures. (c, d) Transformation of (a, b) to linear fitting. (e) Comparison of the intercepts of zero MW field.

As no red shift occurs in the absence of MW irradiation, we infer that the intercept of a single linear fit approximates the ZFS value for zero red shift. Thus, we argue that the intercepts for the two cases should be equal for identical initial thermal equilibrium states of the sample. The practical difference denotes the measurement uncertainty when using the ensemble of $NV^-$ centers presented in our experiment. Accordingly, we compare the intercepts of the fits of the two cases in Fig. 5(e). The differences randomly vary from 29 kHz to 166 kHz over the entire experimental range. The standard deviation of the variation is 49 kHz. Given the normal distribution of the differences, the standard deviation accounts for the standard uncertainty in measuring the ZFS. As stated in the previous section, the red-shift effect is responsible for the large differences of 47-1505 kHz in the measured ZFS values between Case I and Case II. Thus, the asymptotic approximation significantly suppresses the red-shift effect over the entire experimental range.

The authors acknowledge the financial support provided by the Fundamental Research Program of the National Institute of Metrology, China (nos. AKYZD2209-1, AKYRC2303, and AKYRC2301).

Corresponding author:
zhangjint@nim.ac.cn

## Supplemental Material for "Red shift effect on the zero field splitting for negatively charged nitrogen-vacancy centers in diamond"

Wang Zheng [1, 2], Zhang Jintao [2,*], Feng Xiaojuan[2]， Xing Li[2]

[1]*Department of Precision Instruments, Tsinghua University, Beijing, China*
[2] *Division of Thermophysics Metrology, National Institute of Metrology, Beijing, China*

We plotted in Fig. S1 the rises of diamond sample temperatures, together with the ODMR lines, measured from 298.15 K to 323.15 K. As shown for Case I, the sample temperatures get rising at a high rate for 10 s after the MW irradiation, then at a lower rate. The rising level depends on the MW power. The experiments of all the MW powers share the similar variation patterns of sample temperatures. As shown for Case II, a threshold MW field exists. Below the threshold, the sample temperatures get rising at a high rate for 10 s after the MW irradiation. After, the PID controller holds back the sample temperatures around the set value. Beyond the threshold MW field, the controller loses controlling the sample temperatures. For our experiments, the threshold MWs is -5 dBm.

* zhangjint@nim.ac.cn

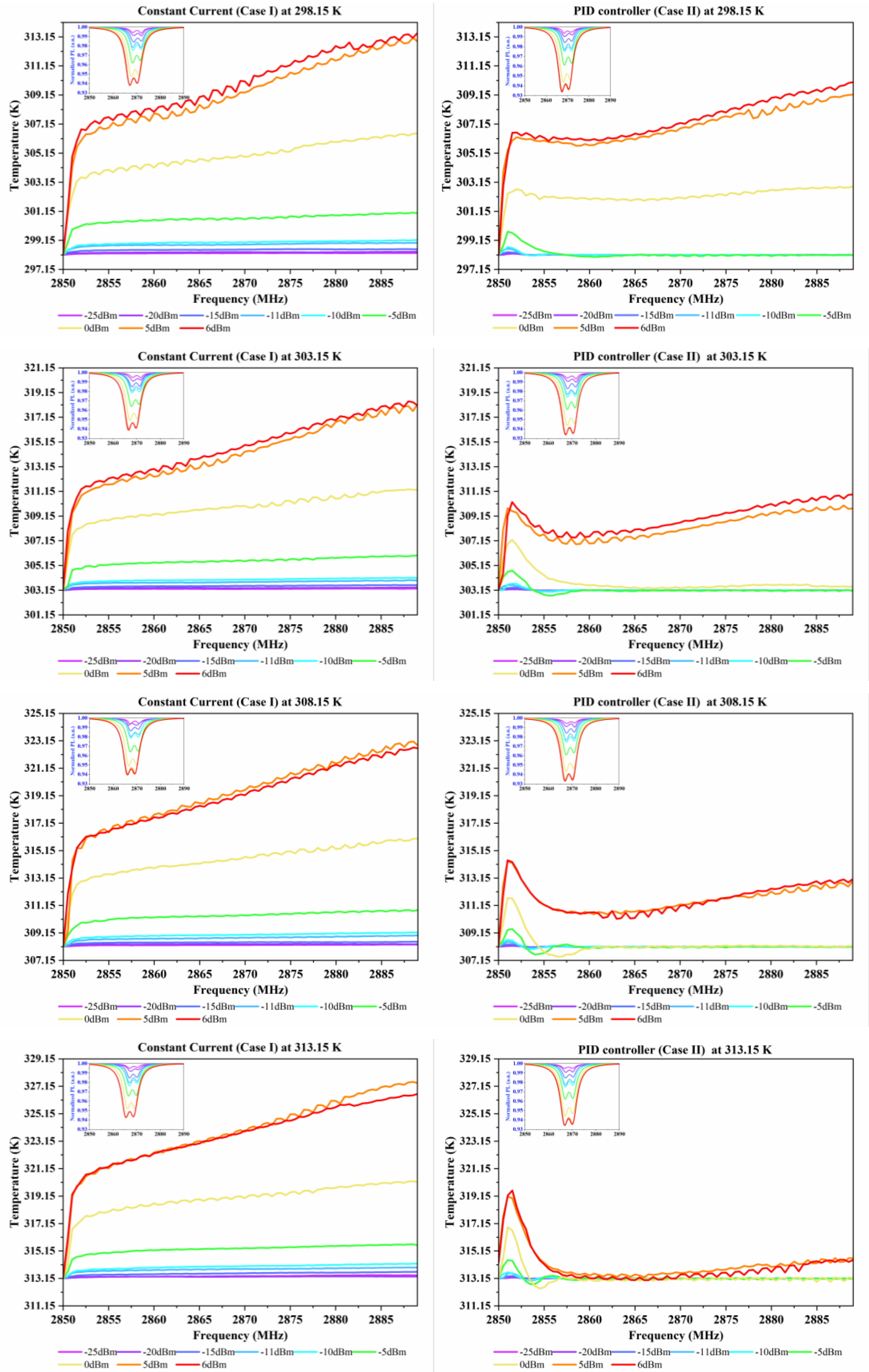

Constant Current (Case I) at 298.15 K
PID controller (Case II) at 298.15 K
Constant Current (Case I) at 303.15 K
PID controller (Case II) at 303.15 K
Constant Current (Case I) at 308.15 K
PID controller (Case II) at 308.15 K
Constant Current (Case I) at 313.15 K
PID controller (Case II) at 313.15 K
Temperature (K)
Frequency (MHz)
Normalized PL (a.u.)
-25dBm -20dBm -15dBm -11dBm -10dBm -5dBm
0dBm 5dBm 6dBm

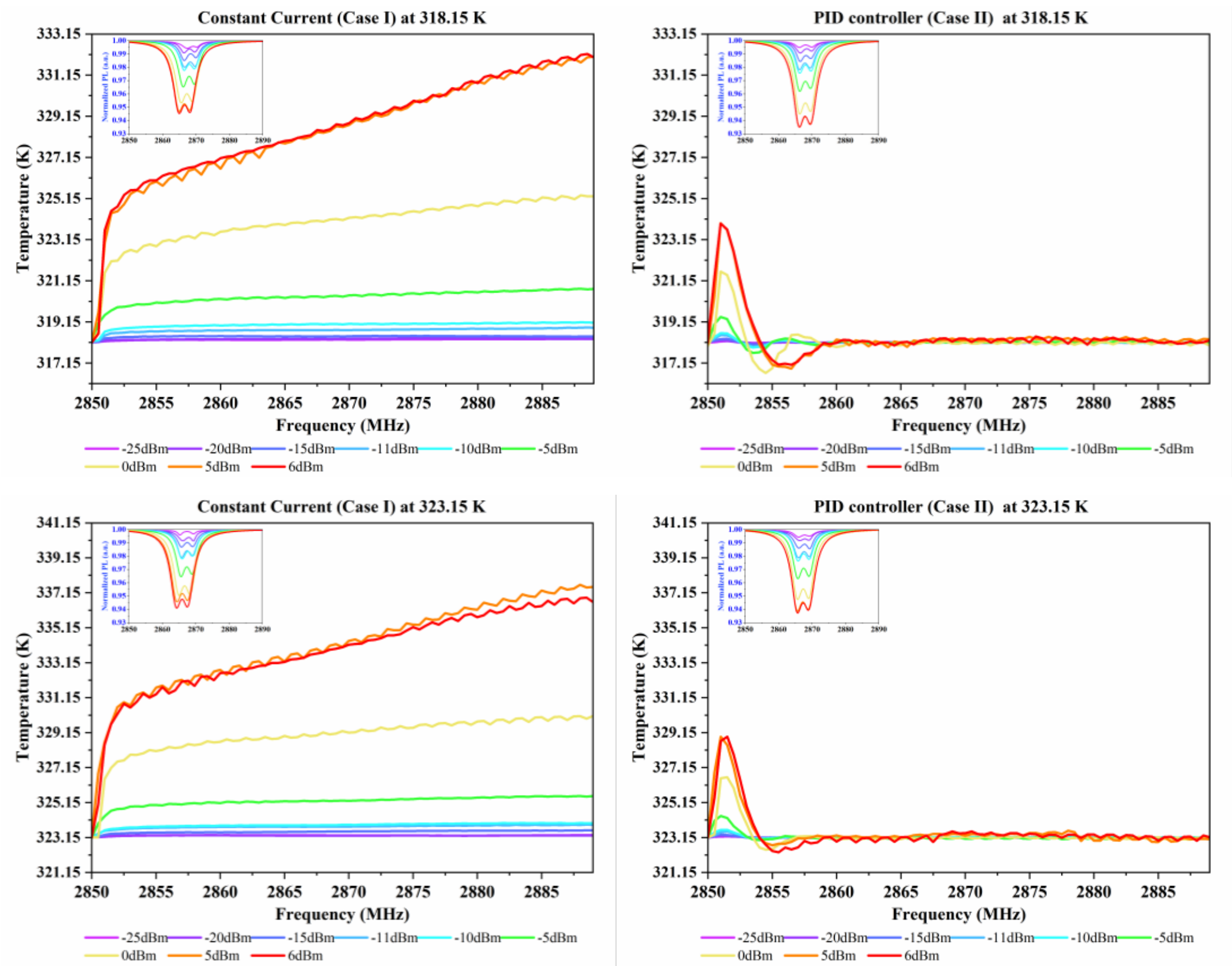

FIG. S1. Plots of sample temperatures and ODMR lines via MW powers